\documentclass[a4paper]{spie}  

\usepackage{amsmath,amsfonts,amssymb,aas_macros}
\usepackage{graphicx}
\usepackage{url}
\DeclareMathOperator{\diag}{diag}
\usepackage[colorlinks=true, allcolors=blue]{hyperref}

\title{BRUTE, PSF Reconstruction for the SOUL pyramid-based Single Conjugate Adaptive Optics facility of the LBT}

\author[a,g,*]{Carmelo Arcidiacono}
\author[a]{Andrea Grazian}
\author[a]{Anita Zanella}
\author[a]{Benedetta Vulcani}
\author[c,a,g]{Elisa Portaluri}
\author[f,g]{Fernando Pedichini}
\author[a]{Marco Gullieuszik}
\author[a]{Matteo Simioni}
\author[f,g]{Roberto Piazzesi}
\author[b,e]{Roland Wagner}
\author[d,g]{Enrico Pinna}
\author[d,g]{Guido Agapito}
\author[d,g]{Fabio Rossi}
\author[d,g]{Cedric Plantet}

\affil[a]{INAF - Osservatorio Astronomico di Padova, Vicolo dell'Osservatorio 5, Padova, Italy, I-35122} 
\affil[b]{Industrial Mathematics Institute, Johannes Kepler University Linz, Altenberger Strasse 69, Linz, Austria, 4040} 
\affil[c]{INAF - Osservatorio Astronomico d'Abruzzo, Via Mentore Maggini, Teramo, Italy, I-64100} 
\affil[d]{INAF - Osservatorio Astrofisico di Arcetri, Via E. Fermi 5, Firenze, Italy, I-50125} 
\affil[e]{RICAM - Johann Radon Institute for Computational and Applied Mathematics, Altenberger Strasse 69, Linz, Austria, 4040}
\affil[f]{INAF - Osservatorio Astronomico di Roma, Via Frascati 33, Monte Porzio Catone, Italy, I-00078} 
\affil[g]{Adoni Laboratorio Nazionale di Astrofisica, Italy} 

\authorinfo{Further author information: (Send correspondence to C.A.)\\C.A.: E-mail: carmelo.arcidiacono@inaf.it, Telephone: +39 049 829 3414}

\begin{document} 
\maketitle

\begin{abstract}
The astronomical applications greatly benefit from the knowledge of the instrument PSF. We describe the PSF Reconstruction algorithm developed for the LBT LUCI instrument assisted by the SOUL SCAO module. The reconstruction procedure considers only synchronous wavefront sensor telemetry data and a few asynchronous calibrations. 
We do not compute the Optical Transfer Function and corresponding filters. We compute instead a temporal series of wavefront maps and for each of these the corresponding instantaneous PSF.
We tested the algorithm both in laboratory arrangement and in the nighttime for different SOUL configurations, adapting it to the guide star magnitudes and seeing conditions. We nick-named it ``BRUTE", Blind Reconstruction Using TElemetry, also recalling the one-to-one approach, one slope-to one instantaneous PSF the algorithm applies.

\end{abstract}


\section{INTRODUCTION}
\label{sec:intro}  
The thirst for enhanced sensitivity brought to extremely sophisticated systems. The current paradigm for optical instrumentation is to design larger and larger telescopes properly armed with Adaptive Optics (AO) modules. Such effort caused an analogous complexity of the camera/spectrograph response. The ground-based PSF, which was seeing dependent, now is time-evolving and spatially varying on minutes and arcsecond scales, respectively. The stochastic behaviour of both the optical turbulence and the telescope vibrations corresponds to an unreliable PSF definition. A long-time average may reduce the divergence with respect to predictions, reducing the problem to an accurate knowledge of the initial conditions. Accurate photometry and morphological studies require accurate PSF knowledge. The PSF information is mandatory, but the availability of point sources is not obvious, since the field of view imaged by AO instruments is typically small (tens of arcsec). 
We focus on the blind PSF-reconstruction\cite{simioni_micado_2020,beltramo-martin_review_2020} (PSF-R). It is “blind” because applies to the cases when the PSF reference is not available: when the only viable way to build a PSF is to use complementary information made available by the observatory (seeing, wind speed) or by the AO instrument (otherwise, i.e., see Massari+~(2020)\cite{massari_successful_2020}). These “telemetry” data offer the opportunity to reduce the PSF object to parameters describing turbulence statistics, AO loop, noise, vibrations, and the optical train. We describe a possible method, we are developing using the SOUL\cite{pinna_soul_2016,agapito_advances_2021}@LBT\cite{hill_large_2010} AO module, and the LUCI\cite{wagner_overview_2014} imager as the reference application.
SOUL recently upgraded the FLAO\cite{FLAO} systems at the LBT replacing the wavefront sensor camera with an OCAM2k\cite{pinna_bringing_2021}.
The FLAO and SOUL are SCAO adaptive optics modules.
Thanks to the low readout noise camera, finer pupil resolution and increased readout frequency SOUL improves several merit figures: sky coverage, magnitude limits, and Strehl Ratio values (optical quality)
SOUL, as the First Light Adaptive Optics (FLAO)\cite{FLAO} did, feeds the NIR camera LUCI.

The authors are collaborating to the development of MORFEO\cite{ciliegi_maory_2018} and MICADO\cite{davies_micado_2010} adaptive optics module and camera for the European ELT\cite{E-ELT}.
MICADO is one of the first light instruments of the ELT. It will be fed by ELT through the MAORY        adaptive optics for Multi-Conjugate Adaptive Optics (MCAO) or through an internal SCAO\cite{clenet_micado-maory_2019} module.
The contribution of the authors to MICADO is in the development of the PSF-R software (MICADO specific).
The important connection is that both MICADO SCAO and SOUL modules use: 
\begin{itemize}
    \item a Pyramid WaveFront Sensor\cite{pyramid} (P-WFS);
    \item a contact-less voice coil deformable mirror controlled in position using capacitive sensors\cite{salinari_study_1994} for optical turbulence correction
\end{itemize} 

\begin{figure}
  \centering
  \includegraphics[width=1.0\linewidth]{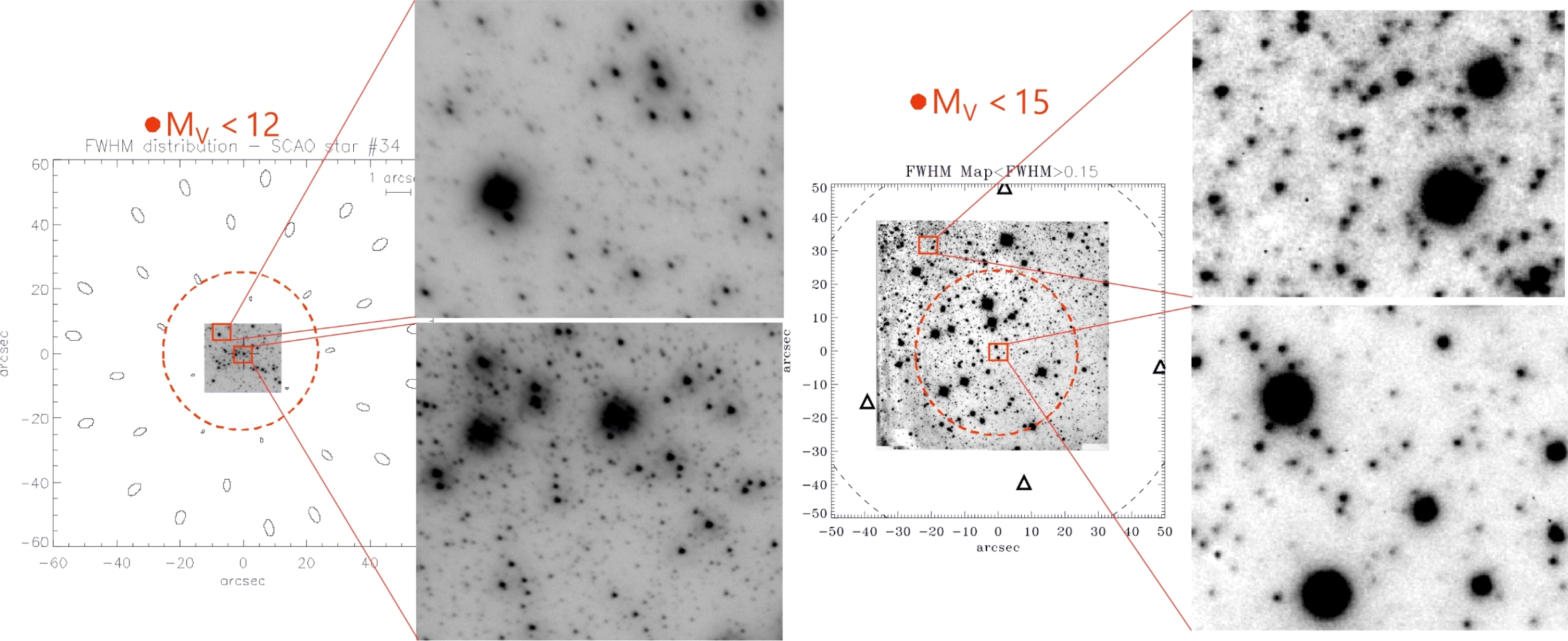}
      \caption {SCAO correction vs MCAO correction. In the MCAO the spatial variability of the PSF is much lower than in the SCAO case. On the left the SCAO M15 core, 23’’x23’’ data J band
PISCES+FLAO @ LBT, inset 6’’x5’’. On the right the MCAO, on the halo of NGC 6388, 65’’x63’’, Ks band, MAD @ VLT, inset 6’’x5’’
}
         \label{fig:SCAO-MCAO}
  \end{figure}

SOUL is the optimal benchmark for the MICADO PSF-R development, see also Simioni+~(2022)\cite{2022arXiv220901563S}. We develop this activity within the MAST\&R working group, and gathered contributions mainly from INAF-Arcetri (SOUL data) and INAF-Padova (MICADO PSF-R team) producing a radial elongation of the PSF centered on the reference star, see Figure~\ref{fig:SCAO-MCAO}.

\section{PSF-R for Single Conjugate Adaptive Optics}
The test case we considered here is the SOUL adaptive optics system for the LBT telescope, feeding the LUCI\cite{seifert_lucifer_2003} Near Infrared camera. In particular, we focus on the on-axis (on the direction of the guide star) case. SOUL is a Single Conjugate Adaptive Optics (SCAO) system. A SCAO modules senses and corrects the turbulence in the direction of the single guide star. and the optical performance peaks in the direction of the guide star. An AO system squeezes the PSF by moving one or more deformable mirrors against the fast optical disturbance injected by the turbulence. 

The SCAO performance ultimately depends on the initial seeing (+telescope vibrations) and on the brightness of the reference star.	
The wavefront correction is limited by loop frequency, actuators spacing (on the deformable mirror), Wave Front Sensor-errors (noise and misalignment) and by the control of the Non-Common Path Aberration (NCPA, instrument versus wavefront sensor).
Adaptive Optics flattens the incoming wavefront and acts moving speckles from the seeing halo to the  aperture mask diffraction pattern.

In the high Strehl Ratio regime (larger than 20-30\%) the PSF is quasi diffraction-limited and present typical structures. The number of modes used to control the deformable mirror fixes the highest spatial frequency corrected:
\begin{itemize}
    \item It is ultimately limited by the number of actuators;
    \item This highest frequency generates a boundary with not corrected one, the so-called control radius\cite{perrin_structure_2003}, well visible in SCAO mode;
    \item Control Radius is at the radius of $\lambda/2d$             where $\lambda$ is the imaging wavelength, and $d$ is the characteristic distance between two adjacent actuators on the deformable mirror. Smaller the distance larger the control radius, at which the halo of the uncorrected component of the seeing  starts.
\end{itemize}

\section{PSF-R blind method ``BRUTE - Blind Reconstruction Using TElemetry"}  
\subsection{BRUTE PSF-R in short}  
To distinguish this method from other blind methods using only adaptive optics system and telescope telemetry data\cite{wagner_point_2018,veran_estimation_1997,fusco_reconstruction_2020,ragland_point_2016}, we nick-named this BRUTE, Blind Reconstruction Using TElemetry, also recalling the one-to-one method, one slope-to one instantaneous PSF, we follow. We build a wavefront array for each iteration of the adaptive optics loop that was saved. We use the signal from the wavefront sensor, in our case strictly related to the XY gradient of the wavefront, to generate a wavefront map, and we add to this that part of the wavefront unseen by the wavefront sensor.
BRUTE uses the adaptive optics loop information registered simultaneously with the science frame.
BRUTE fits turbulence on the pseudo open loop data and noise models on the slopes statistics. Pseudo-open loop data is the description of the original open loop (wavefront without adaptive correction) made by summing synchronous wavefront slopes contributions and deformable mirror shape.
It applies an instrumental look up table from calibrated P-WFS optical gain curves\cite{korkiakoski_improving_2008,deo_modal_2018} finally modulated for the best estimation of its amplitude. 
The algorithm applies calibration made for each of the operative modes of the SOUL instrument, which adapts the pupil resolution (CCD binning) and camera read-out frequency to the reference star brightness.
For each wavefront, BRUTE computes the instantaneous PSF corresponding to the modeled wavefront map and considers the actual pupil mask definition.
In the following sections, we take as a reference the case of SOUL at LBT. It to be stressed that in the case of SOUL the optical gain is monitored during the operation and applied to properly perform the NCPA suppression\cite{esposito_-sky_2020}. 
In the algorithm, regardless of the origin of the residuals, all the wavefronts components are described as a  combination of shapes (typically mirror modes, or Zernike modes). 
\subsection{What the telemetry data are}    
The BRUTE PSF-R algorithm requests data synchronous to the observations. This condition allows accounting for specific incident happening during open shutter time (wind gusts, seeing spikes, and una-tantum vibrations and loop instabilities). 

{\em Wavefront Sensor Slopes:} the camera-synchronous wavefront sensor measurements history: it can be a large data sets since grows at loop frequency. Each iteration step is large as twice the number of active sub-apertures.

{\em Deformable Mirror History:} the camera-synchronous deformable mirror shape history: growing at loop frequency and large as the number of controlled modes. The control loop describes the deformable mirror shape as linear a combination of the i$^{th}$-shapes composing the deformable mirror modal description.

{\em Loop Gain:} if applied, it is a vector of modulation amplitude,  long as the number of modes in use, that defines for each mirror mode the fraction of it to be added to the deformable mirror.

Better is the coverage of the observation with the registration of these data, higher is the precision of the reconstructed PSF. To these quantities that varies during the operation, others calibration or pre-processed data are needed for the blind PSF-R.

{\em IM:} it is the deformable mirror to wavefront sensor interaction matrix, this and its the pseudo-inverse $IM^{-1}$ will be used by the BRUTE algorithm

{\em R:} is the control matrix, this is known and determined using the calibrated interaction matrix ($IM$);

{\em KL:} is the array containing the wavefront map of each of the modes in the modal base (KL stands for Karhounen-Loeve\cite{karhunen_uber_1947}).

{\em NCPA} is the vector of modal coefficients containing the modal description of the wavefront difference between wavefront sensor and scientific camera.
\subsection{Optical Loop Gain}
Given the information in the telemetry, ideally, the instantaneous wavefront seen by the wavefront sensor is: \[{WF}_{SLP} = slopes \times IM^{-1} \times KL \] In a closed loop system, this ${WF}_{SLP}$ is a low-resolution image of the original residual wavefront incident on the wavefront sensor. It is the so-called “parallel part“ mostly responsible for the shape of the PSF within the control radius.
The parallel part of the wavefront, the ${WF}_{SLP}$, history is then computed from the slopes records.
However, several corrections need to be implemented, as for the noise component and for the optical loop gain $ g_{opt}$ mentioned above.
The proper derivation of the wavefront map from the slopes requires the correction of the non-linearity effects. These makes deviating the response amplitude of the wavefront sensor to the proportional relation with incident wavefront amplitude. The deviation depending also on the spatial scale of the measured mode.
       \[{WF}_{SLP} = slopes \times  {IM}^{-1} \cdot {g}_{opt} \times  KL\]

In the P-WFS case the optical loop gain, should be calibrated using dedicated strategies, see Agapito+~(2021)\cite{agapito_advances_2021} or Deo+~(2022)\cite{deo:hal-03084867} for application to optimal loop control.

\subsection{Noise part into the parallel component}

From the history of the slopes we can derive both noise and turbulence contributions. 
The noise component is part of the parallel wavefront. It is added to the estimated wavefront as part of the $WF_{SLP}$.
We derive the modal coefficients vector of the parallel part of the wavefront, $m$, from the slopes: 
$m = slopes \times  {IM}^{-1} \cdot {g}_{opt}$ returns the elements of the linear combination of wavefront shapes composing the instantaneous $WF_{SLP}$.
The amplitude corresponding to the noise part in the $m$ is removed, by scaling the $m$ mode by mode following the $\diag{{IM}\,{IM}^T}$, and the remaining part is associated to the optical turbulence and transferred as is into the $WF_{SLP}$..
\subsection{Orthogonal component and derived quantities}
The BRUTE PSF-R algorithm estimates the characteristics length (r$_0$) of the spatial power spectral density of the turbulence, fitting an analytical model onto the deformable mirror history and wavefront sensor data.
We used here a modal representation of the power spectral density (PSD) adapting the Noll~(1976)\cite{noll_zernike_1976} formula, fitting the power law on the power spectral density of the modal description of pseudo open loop WF.
The point is to identify in that PSD the original atmospheric turbulence contribution, separating the effect of the noise.
We assumed that tip-tilt modes are dominated by the telescope vibrations, thus excluding these from the power law fit.
The fit allows extrapolating the modal amplitude of the uncontrolled components (the modes- or the frequencies higher than the one controlled by the SCAO system), finally estimating the orthogonal part of the wavefront.

I.e. in SOUL data, we had 500 KL modes, and the orthogonal components extend up to Zernike 3000.
\subsection{Building the PSF}
\begin{figure}
  \centering
  \includegraphics[width=1.0\linewidth]{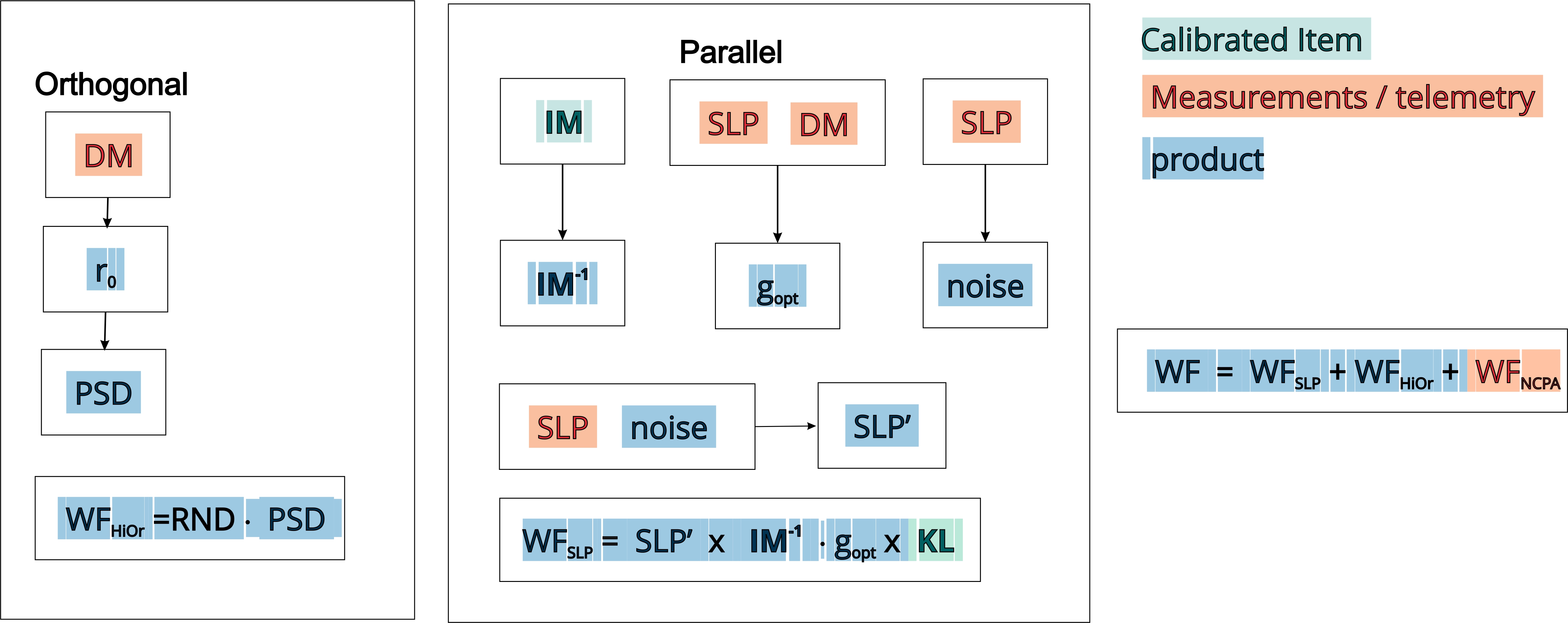}
      \caption {Scheme dividing parallel and orthogonal components. Different colours indicate if the quantities in use are calibrated, measured or secondary products.
}
         \label{fig:BRUTE-scheme}
  \end{figure}

The processes above are listed schematically in Figure~\ref{fig:BRUTE-scheme}
The integrated PSF is the sum of a number of realizations, one for each $k$-slope vector frame:
\[
{{WF}_{rad}}_{k} = 2\pi \cdot {WF}_{SLP} + {WF}_{HiOrder} + {WF}_{NCPA} \\
\]
\[
PSF = {\sum}_k {\mid pupil \cdot e^{i{{{WF}_{rad}}_{k}}} \mid}^2
\]

\section{BRUTE PSF-R method for SOUL-LUCI}
\subsection{Optical loop gain}
\begin{figure}
  \centering
  \includegraphics[width=0.66\linewidth]{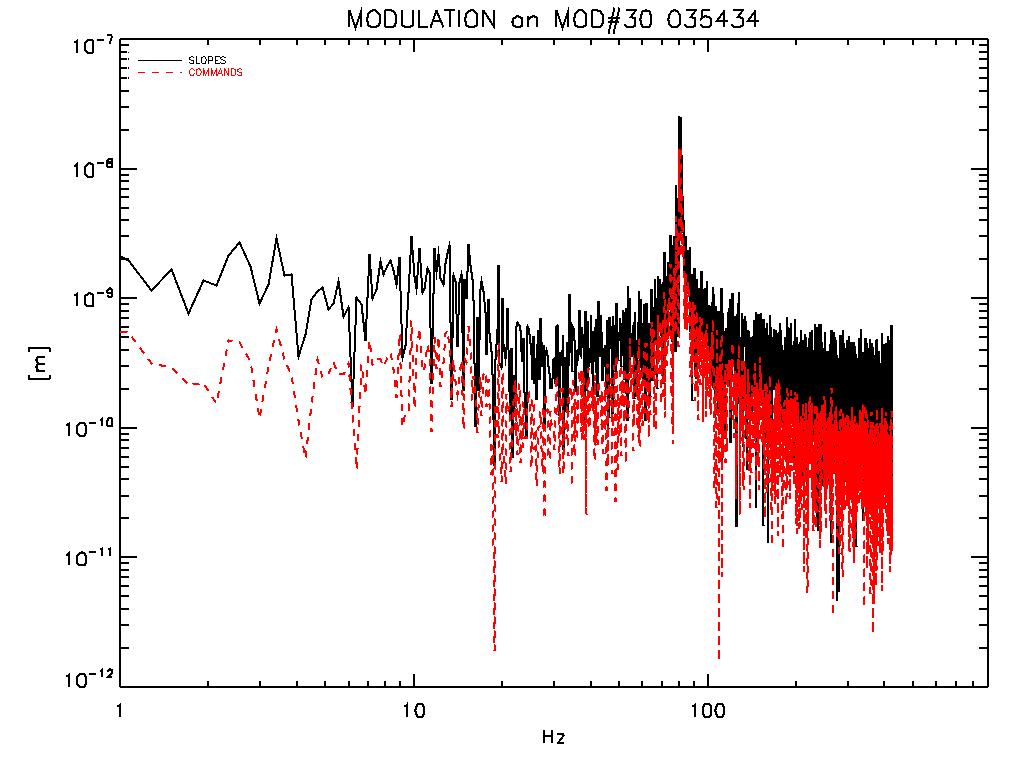}
      \caption {Temporal power spectral density of the slopes response (black, solid) and of the deformable mirror command for modulated mode \# 30,
}
         \label{fig:modulation30}
  \end{figure}

SOUL continuously tracks the optical gain (sensitivity), by measuring the feedback to a sinusoidal modulation injected on a single mode (\#30), Figure~\ref{fig:modulation30}.
The slopes history saved are already corrected for this value. However, this estimation can be refined in the post-processing, 
and the algorithm re-computes it using again the peak of temporal power spectrum corresponding to the  continuous modulation of KL mode \#30, and now comparing the ratio between the amplitudes of modes \#30 as modulated by the deformable mirror over the  modes \#30 as derived by the wavefront sensor with respect to similar ratio on adjacent modes.
This correction is applied mode by mode scaling an empirical relation calibrated using ``daytime" data, in which the optical distortion evolution is known, see also Deo+~2018\cite{deo_modal_2018}.

\subsection{Pupil Masks-Cold stop and Parallactic angle}
\begin{figure}
  \centering
  \includegraphics[width=1.0\linewidth]{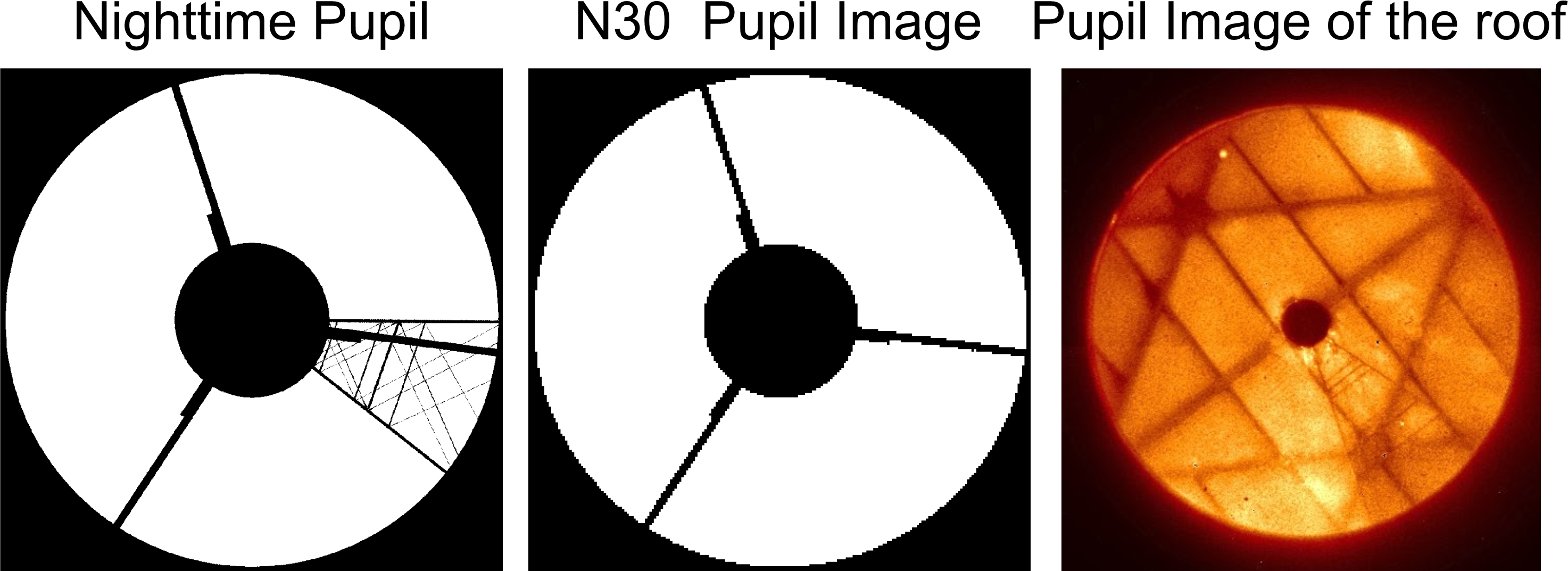}
      \caption {From left to right. On-sky pupil definition,the LUCI N30 pupil and an image taken in pupil camera mode (not the N30) during the alignment of LUCI. We used this kind of image to calibrate pupil rotation angle with the LUCI rotator angle movement.
}
         \label{fig:pupil}
  \end{figure}

LUCI foresees three different level of zoom. The one designed for Adaptive Optics application is the N30 camera\cite{heidt_commissioning_2018}.
The N30 camera is a reversed telescope design that includes its ``secondary mirror" inserted in the optical axis and mounted on three arms: the diffraction pattern due to these arms follows the apparent sky rotation, However, the effective pupil illumination of LUCI depends whether working on daytime (laboratory) or nighttime (on-sky), since in the laboratory configuration ``daytime"\cite{2008SPIE.7015E..27R} LUCI does not see the M2 spider arms (and the primary mirror).
Actual size of the telescope is 8.06 meter because of the extended cold stop in LUCI, and central obstruction is 0.314, because of the obstruction imposed by the secondary mirror of the reflective design of the N30 camera, see Figure~\ref{fig:pupil}.

\subsection{Pupil Masks, cold stop and parallactic angle}

Daytime pattern does not rotate, while the M2-M3 spider arms rotate with the parallactic angle as seen from detector.
To build the nighttime on-sky pupil we need to rotate the spider pattern mask for the corresponding parallactic angle and add this to the N30 pupil mask.
\subsection{Filter width and Atmospheric Dispersion}
The PSF resulting from previous steps are monochromatic. The smoothing effect due to the wavelength dispersion along the width of filter is then applied, considering the slight changes of the PSF shape across the filter.

\begin{figure}
  \centering
  \includegraphics[width=0.66\linewidth]{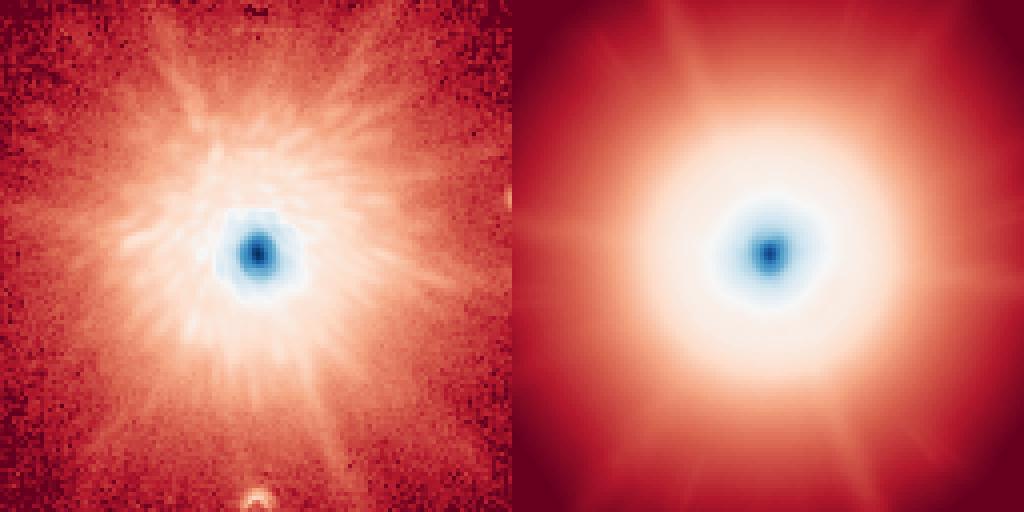}\,\includegraphics[width=0.33\linewidth]{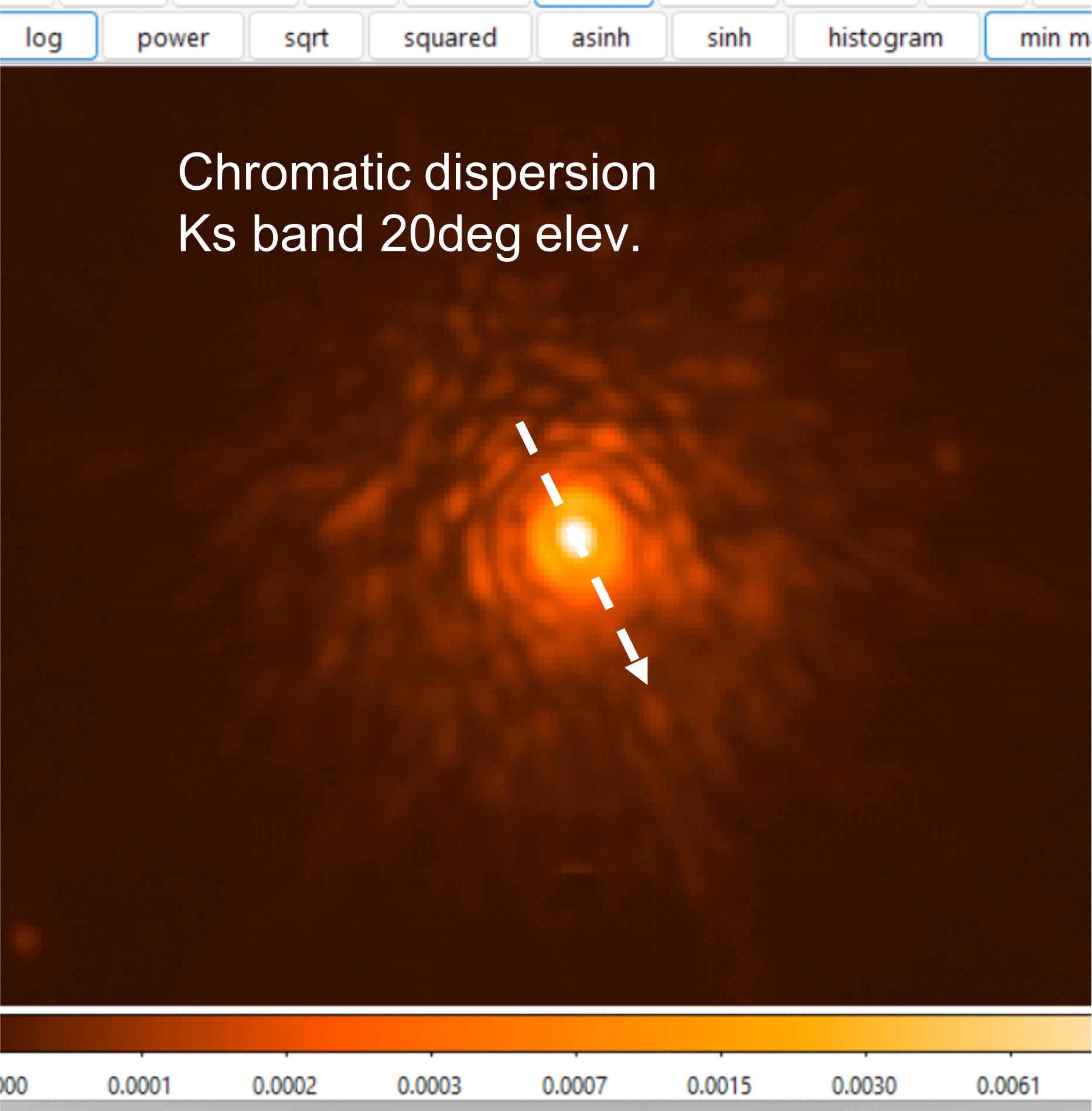}
      \caption {On the left an image taken with LUCI using BIN 2 mode (200modes). The filter is the broadband J. On the central panel the PSF reconstructed using BRUTE. On the right the direction of dispersion is made evident by the arrow. 
}
         \label{fig:chromatic-dispersion}
\end{figure}

The LUCI camera does not mount any atmospheric dispersion compensator. As a result, the diffraction-limited images are elongated along the telescope elevation axis direction, Figure~\ref{fig:chromatic-dispersion}. This elongation depends mainly on the wavelength, the filter width and on the elevation angle, and also on other parameters that were not tracked during the observation we used (humidity, pressure vertical profile) or have a negligible effect as the color of the star (we observed in the Near Infrared: the spectrum of non-peculiar stars is quite flat across the wavelengths in the broad-band filters).
So, the last step for the BRUTE PSF-R is the application of a correction for wavelengths dependant effects on the integrated PSF computed above.
We use the analytic formula that gives the chromatic dispersion with respect to the elevation, applying small shift of the monochromatic PSFs along the dispersion direction.
\section{BRUTE PSF-R application to SOUL-LUCI data}
\begin{figure}
  \centering
  \includegraphics[width=1.0\linewidth]{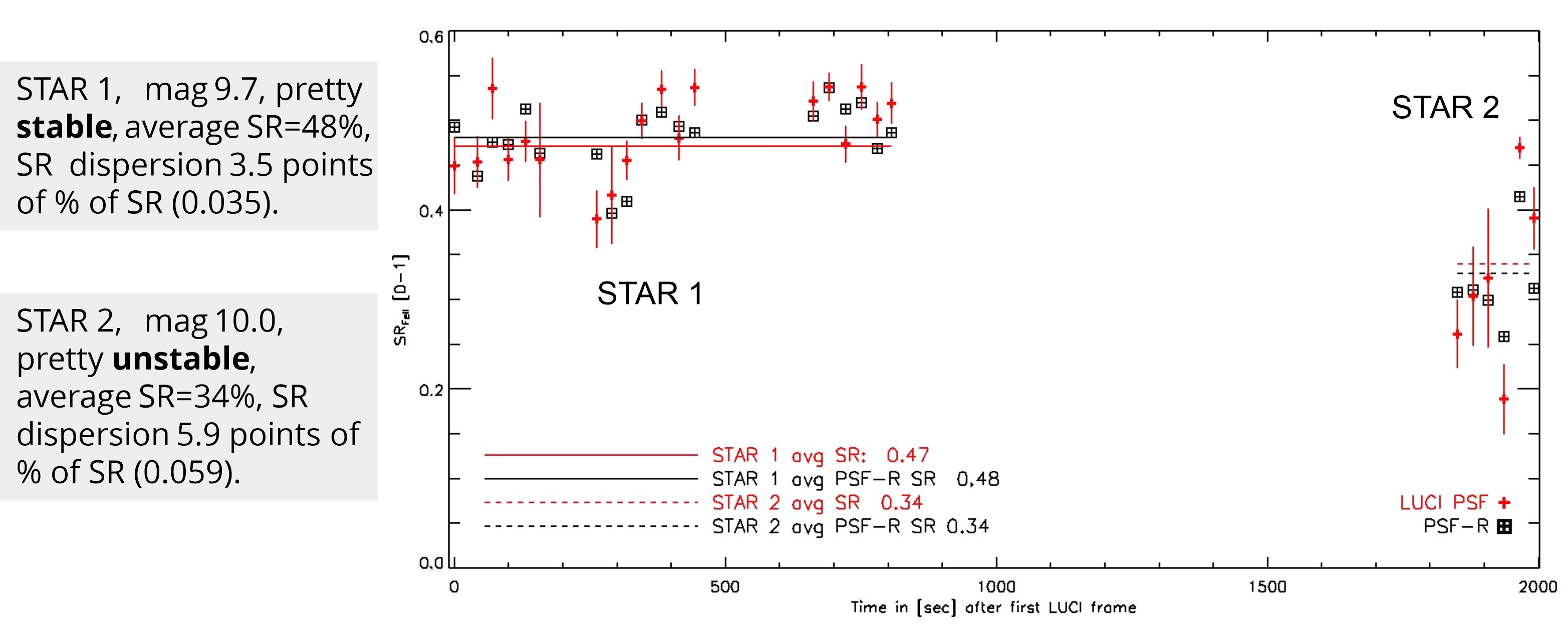}\
      \caption{25 LUCI observations, each one simultaneously registered with SOUL wavefront sensor telemetry on the LBT left side, two different targets, with similar magnitudes. See insets on the left of the graph.
}
         \label{fig:bin1}
\end{figure}

We applied the method on 25 LUCI SOUL frames acquired during commissioning time in a single night of observation. Each LUCI frame was simultaneously registered with SOUL wavefront sensor telemetry. For that commissioning run (11-2019) only the LBT left side was available. We had two different targets, with similar magnitudes. We observed 25 sets of 0.3sec $\times$ 20 images in the narrow filter FeII (1.645 micron). See Figure~\ref{fig:bin1}. 
The first star presents a quite stable loop, modulated by the seeing variation, while the second had loop instabilities that make an extreme variation of the performance frame by frame.
Despite the two having different behavior the BRUTE algorithm can nicely match the average Strehl Ratio and Encircled Energy profile with errors in the stable case is limited to a few points of percentage in line with the dispersion of the original Strehl Ratio measurements itself.
\begin{figure}
  \centering
  \includegraphics[width=0.8\linewidth]{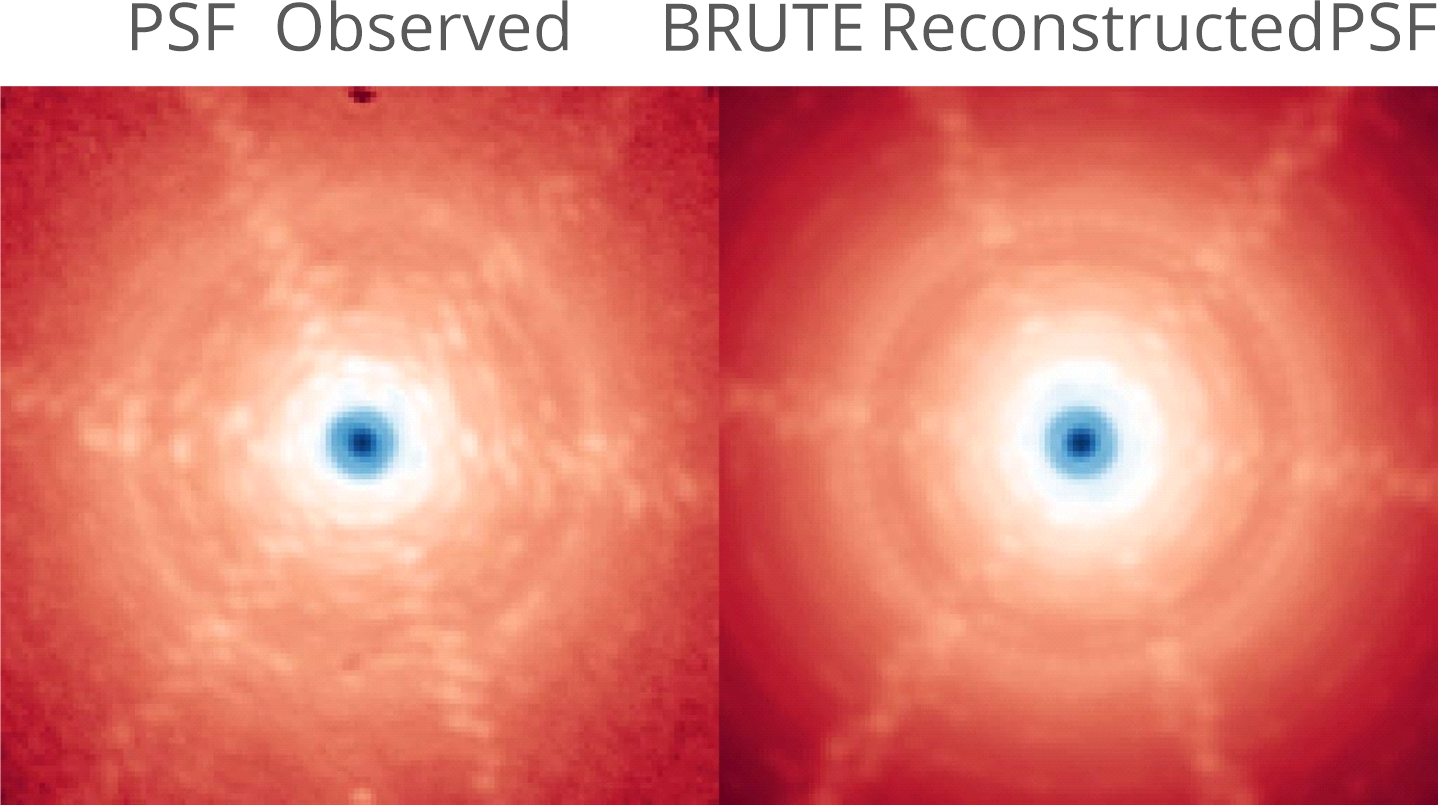}
      \caption {On the left the average image of the 25 frames of the observed SOUL@LUCI PSF and on the right the mean of the BRUTE PSF-R PSF reconstructions. The mean of the 25 frames gives the possibility to compare also the low signal noise ratio halo of the PSF, otherwise marginally visible in the original LUCI frame.
}
         \label{fig:bin1-PSF}
  \end{figure}
\section{Conclusions}
We reported about the successful application of the BRUTE blind PSF reconstruction method to on sky LUCI@LBT data. The SOUL adaptive optics system offers the telemetry data necessary to extract the wavefront components: open loop turbulence, noise and pyramid optical
gain. We successfully matched the high Strehl Ratio PSF and demonstrated the method valid on high
Strehl Ratio conditions. We tested a-posteriori correction for the pyramid wavefront sensor optical gain. 
The results obtained on the LUCI@SOUL data are very promising for MICADO PSF-R development, and the BRUTE PSF-R system can be offered right now as support for the observations (on our server).
We use the history of the deformable mirror shape for the estimation of the turbulence PSD.

However, this opportunity is less valid in the case that
a few modes are controlled, 
and/or
correction is poor (the optical surface of the deformable mirror is not half of the negative copy of the reference star WF).

Next step is to prove the method on lower Strehl Ratio observations characterized by a larger noise component.

\acknowledgments 
The  work has been supported by INAF through the Math, ASTronomy and Research (MAST\&R), a working group for mathematical methods for high-resolution imaging.

\bibliography{report}

\bibliographystyle{spiebib} 

\end{document}